\documentstyle[12pt]{article}
\newcommand{\be}{\begin{equation}}
\newcommand{\bea}{\begin{eqnarray}}
\newcommand{\eea}{\end{eqnarray}}
\newcommand{\ba}{\begin{array}}
\newcommand{\ea}{\end{array}}
\newcommand{\ee}{\end{equation}}

\expandafter\ifx\csname mathbbm\endcsname\relax

\else

\fi \textheight 22cm \textwidth 15cm \topmargin 1mm \oddsidemargin
5mm \evensidemargin 5mm

\newcommand{\nn}{\nonumber}
\newcommand\para{\paragraph{}}
\newcommand{\ft}[2]{{\textstyle\frac{#1}{#2}}}
\newcommand{\eqn}[1]{(\ref{#1})}

\def\Dslash{\,\,{\raise.15ex\hbox{/}\mkern-12mu D}}
\def\Dbarslash{\,\,{\raise.15ex\hbox{/}\mkern-12mu {\bar D}}}
\def\delslash{\,\,{\raise.15ex\hbox{/}\mkern-9mu \partial}}
\def\delbarslash{\,\,{\raise.15ex\hbox{/}\mkern-9mu {\bar\partial}}}
\def\pslash{\,\,{\raise.15ex\hbox{/}\mkern-9mu p}}
\def\calDslash{\,\,{\raise.15ex\hbox{/}\mkern-12mu {\cal D}}}

\input amssym.def
\input amssym.tex

\def\bbox{{\,\lower0.9pt\vbox{\hrule \hbox{\vrule height 0.2 cm
\hskip 0.2 cm \vrule height 0.2 cm}\hrule}\,}}

\def\Dslash{\,\,{\raise.15ex\hbox{/}\mkern-12mu D}}
\def\Dbarslash{\,\,{\raise.15ex\hbox{/}\mkern-12mu {\bar D}}}
\def\delslash{\,\,{\raise.15ex\hbox{/}\mkern-9mu \partial}}
\def\delbarslash{\,\,{\raise.15ex\hbox{/}\mkern-9mu {\bar\partial}}}
\def\pslash{\,\,{\raise.15ex\hbox{/}\mkern-9mu p}}
\def\calDslash{\,\,{\raise.15ex\hbox{/}\mkern-12mu {\cal D}}}

\begin{document}
\pagestyle{plain} \setcounter{page}{1}
\newcounter{bean}
\baselineskip16pt

\begin{titlepage}

\begin{center}
{\today}
\hfill {IPM/P-2004/016 \\
\hfill MIT-CTP-3488 \\
\hfill SLAC-PUB-10406\\
\hfill SU-ITP-4/15 \\}

\vskip 1cm {\Large \bf DBI in the Sky \\} \vskip 1.3cm

{Mohsen Alishahiha$^{1}$, Eva Silverstein$^{2}$ and David Tong$^{3,2}$}\\

\vskip 1.3 cm

{\sl ${}^1$ Institute for Studies in Theoretical Physics
and Mathematics (IPM)\\
P.O. Box 19395-5531, Tehran, Iran \\
{\tt alishah@ipm.ir}}

\vskip .2cm
{\sl ${}^2$SLAC and Department of Physics, Stanford University, \\
Stanford, CA 94309/94305, U.S.A. \\
{\tt evas@slac.stanford.edu} }

\vskip .2cm
{\sl ${}^3$ Center for Theoretical Physics,
Massachusetts Institute of Technology,
\\ Cambridge, MA 02139, U.S.A. \\
{\tt dtong@mit.edu}}

\end{center}

\vskip 0.2 cm
\begin{abstract}

We analyze the spectrum of density perturbations generated in models of the recently discovered ``D-cceleration"
mechanism of inflation. In this scenario, strong coupling quantum field theoretic effects sum to provide a
DBI-like action for the inflaton. We show that the model has a strict lower bound on the non-Gaussianity of the
CMBR power spectrum at an observable level, and is thus falsifiable.  This in particular observationally
distinguishes this mechanism from traditional slow roll inflation generated by weakly interacting scalar fields.
The model also favors a large observable tensor component to the CMBR spectrum.


\end{abstract}

\end{titlepage}

\section{Introduction and Summary}

Inflationary model building is a well developed subject, but has largely focused on weakly coupled scalar fields
rolling around in a potential.  The small parameters in the potential, which are required for slow roll
inflation with the right magnitude of density perturbations, lead to primordial density perturbations which are
Gaussian to a very high degree of accuracy \cite{Komatsu:2003fd}\cite{Acquaviva:2002ud}\cite{juan}.  In this
paper we will show that the set of strongy coupled field theoretic models of inflation introduced in
\cite{Dccel}\ produces distinctive predictions for the CMBR spectrum, including a much larger observable level
of non-Gaussianity and a strong preference for large observable tensor modes. This renders the model
observationally distinctive and falsifiable.



\para
In \cite{Dccel} a ``D-cceleration'' mechanism for
slowing scalar field motion was identified in the context of strongly 't Hooft
coupled conformal field theories. The slow motion of the
scalars is understood in a simple way from the gravity side of the AdS/CFT correspondence which provides the
effective description of the system. The inflaton field $\phi$ is a scalar parameterizing a direction on
the approximate Coulomb branch of the system; motion toward the origin is motion of a 3-brane domain wall in
the radial direction of $AdS_5$, toward the horizon where new degrees of freedom $\chi$ become
light. The motion is constrained by the causal
speed limit on the gravity side, leading to slow roll even in the presence of a relatively steep potential.
In terms of the effective action this behavior occurs due to the higher derivative terms
encoded in the Dirac-Born-Infeld
(DBI) action which arise from integrating out $\chi$ fields\footnote{At strong 't Hooft coupling
these virtual effects  dominate over particle production of $\chi$s. Related ideas using particle production --
which dominates at weak coupling -- to
slow down the field appear in \cite{trappedinflation}. Earlier studies of the dynamics of such probes appear in
\cite{juanDBI,KL}.}.  When the effective field theory is subsequently coupled to gravity, this mechanism leads
to inflation in much steeper potentials than are allowed in standard weakly coupled slow roll inflation.

\para
In this paper, we will analyze the density perturbations resulting from a simple set of inflationary models
based on the D-cceleration mechanism proposed in \cite{Dccel}.  In order to suppress the magnitude of the power
spectrum to observed levels, we must introduce a large number (as in all known models of inflation):  the radius
of curvature on the gravity side must be at least of order 100 in string units.  Given this, we find a
parameteric and numeric lower bound on the non-Gaussianity in the spectrum at observable levels.  In order to
avoid overly large non-Gaussianity as inflation progresses, we must start with field VEVs of order the Planck
scale. The observational upper bound on non-Gaussianity, when combined with field VEVs not much larger than the
Planck scale, leads to a lower bound on the tensor component to the spectrum, also at observable levels. These
features together provide a clean empirical test distinguishing this class of models from all others known to
us.

\para
The origin of the strong non-Gaussianities can be understood rather simply.  The DBI Lagrangian is proportional
to $\sqrt{1-v_p^2}$, where $v_p=\sqrt{\lambda}\dot\phi/\phi^2$ (for 't Hooft coupling $\lambda$) is the
gravity-side proper velocity of the brane probe whose position collective coordinate is the inflaton $\phi$. The
inflationary solution in \cite{Dccel} involves a proper velocity approaching the speed of light as $\phi$
approaches the origin (with both $\phi$ and $\dot\phi$ decreasing toward zero).  Expanding the action in
fluctuations of $\phi$ involves expanding the square root in the DBI Lagrangian, which produces powers of
$\gamma=1/\sqrt{1-v_p^2}$ accompanying powers of the fluctuations of the inflaton.  Since $\gamma$ is relatively
large, this produces a large contribution to non-Gaussianities in the model.

\para
Although the context of our inflationary mechanism is not particularly exotic -- the system consists of quantum
field theory coupled to gravity, with strong quantum effects in the field theory playing an important role in
the dynamics -- there are nonetheless several novelties in our calculation compared with the usual story of
density perturbations \cite{mfb,ll}. We will see that fluctuations of $\phi$ travel at a small sound speed
$c_s\ll 1$. In systems with such a sound speed, the fluctuations freeze out not at the Hubble horizon $H^{-1}$,
but at the smaller ``sound horizon'' $c_sH^{-1}$. Moreover, for sufficiently large momenta, the fluctuations of
the inflaton $\phi$ couple to the full relativistic conformal field theory; this turns out to be a subleading
effect in our dynamics.


\para
Some aspects of our analysis have antecedents. Higher derivative contributions to the Gaussian spectrum of
density perturbations were studied model independently in \cite{GM} (where they were applied to the case of
k-inflation \cite{damour}) while a particular higher derivative contribution to non-Gaussianity was studied in
\cite{Creminelli:2003iq}.  In our case, the higher derivative corrections are crucial to the dynamics near the
origin since they are only suppressed by powers of the $\chi$ mass rather than by a fixed high UV mass scale as
in \cite{Creminelli:2003iq}.  Recently an interesting proposal for a new type of inflationary matter sector
involving ghost modes stabilized by higher derivatives appeared in \cite{Arkani-Hamed:2003uz}; the magnitude of
non-Gaussianities arising in ghost inflation is similar to that of our strongly coupled field theory, though in
our case we favor observable tensor modes while in the ghost case these are predicted to be negligible.  More
generally, as articulated clearly in \cite{Creminelli:2003iq,Arkani-Hamed:2003uz,juan}, the detailed angular
dependence of the non-Gaussianity offers a way to differentiate between the D-cceleration mechanism, ghost
inflation, and other non-trivial interactions occurring outside the horizon such as those described in
\cite{Lyth:2002my,Dvali:2003em,Kofman:2003nx,Zaldarriaga:2003my,Bernardeau:2002jy}.


\para
In solving the various constraints from observational data, we find that our model must satisfy several strong
constraints on its parameters which we should emphasize.  One such constraint, as mentioned above, is a large
curvature radius of the warped throat.  Another is that the warped compactification forming our background
geometry must be somewhat anisotropic.  Finally, in order to avoid large non-Gaussianity by the end of our
inflationary window, the VEV of our inflaton must be of the order of the Planck scale (depending somewhat on the
details of the warp factor chosen and potentially requiring a functional fine tune as in the earliest models of
weakly coupled inflation). In any case, the strong level of predictivity in the model leaves us able to let the
data decide its ultimate relevance.

\para
The paper is organised as follows. In Section 2, we briefly review the model of \cite{Dccel} and describe some
velocity dependent effects pertaining to our background of a rolling scalar field. In Section 3 we study the
spectrum of Gaussian perturbations in our model, deriving expressions for the normalization and tilt of the
scalar and tensor modes. In Section 4 we turn to the non-Gaussianities in our model and compute the fluctuation
three-point function. In Section 5, we compare expressions to the CMBR data, fixing our parameters and deriving
predictions for future experiments. We also include a discussion about the amount of fine-tuning required. We
conclude in Section 6 with a discussion of our results and some future directions. We include two appendices:
Appendix A contains a derivation of the field-velocity corrections to particle masses and couplings which plays
some role in our analysis and which has wider applicability to systems of rolling scalar fields on branes, while
Appendix B collects formulae for the CMBR spectrum as a function of general warp factor and potential.

\section{D-cceleration: A Review}

In \cite{Dccel} a new mechanism for slow roll inflation was introduced, motivated by the behavior of rolling
scalar fields in strongly interacting theories, analyzed using the AdS/CFT correspondence. The key feature of
this model is that the inflaton field $\phi$ is naturally slowed as it approaches a point where many light
degrees of freedom $\chi$ emerge, the slowdown arising from the virtual effects of the light particles. From the
bulk perspective, this scenario translates to a probe D3-brane travelling down a five-dimensional warped throat
geometry (with some internal space $X$, such as the $S^5$ of the simplest version, completing the solution in a
model dependent way). The UV end of the throat joins smoothly onto a compactification in the manner of
\cite{warped}\cite{Randall:1999ee}, ensuring that gravity in four dimensions is dynamical while coupling the
field theory to sectors in the compactification which can generate corrections to the effective action for
$\phi$. These corrections generically produce a nontrivial potential energy for $\phi$ including a mass term
$m^2\phi^2$ and a corresponding closing up of the throat in the IR region at a scale $\phi_{IR}\sim
\sqrt{g_s}m$. The whole set-up is similar to that considered in \cite{kklmmt} but the inflationary process takes
place in a somewhat different regime of parameter space: with a steep potential and a fast moving D3-brane.

\para
The dynamics of the probe D3-brane is captured by the Dirac-Born-Infeld (DBI) action
coupled to gravity,
\be
S = \int \ft12 M_p^2 \sqrt{-g} {\cal R} + {\cal L}_{\rm eff} +\ldots
\label{act1}\ee
with
\be
{\cal L}_{\rm eff} = -\frac{1}{g_{s}}\sqrt{-g}\left(\, f(\phi)^{-1}
\sqrt{1+f(\phi)g^{\mu\nu}\partial_\mu\phi\partial_\nu\phi}+V(\phi)\right)
\label{act2}\ee
Here we work with the reduced Planck mass $M_p\approx 2.4\times 10^{18}$ GeV. The function $f(\phi)$ is the
(squared) warp factor of the AdS-like throat. For example, for a pure $AdS_5$ of radius $R$, it is simply
$f(\phi)= \lambda/\phi^4$ with $\lambda\equiv R^4/\alpha^{\prime\,2}$. In the $SU(N)$ ${\cal N}=4$ super Yang
Mills theory, $\lambda=g_s N$ where $g_s$ is the squared gauge coupling; in more general models such as
orbifolds, $\lambda$ will depend on a more complicated way on the brane numbers and other integer quantum
numbers. As mentioned above, the throats arising from IIB flux compactifications are not AdS at all length
scales, but nonetheless can look approximately AdS in some window, so that \be f(\phi)\approx
\frac{\lambda}{\phi^4}\ \ \ \ \ {\rm for}\ \phi\in (\phi_{IR},\phi_{UV}) \label{f}\ee where $\phi_{UV}\leq
R/\alpha^\prime$, the scale at which the throat matches onto the full compactification geometry.  For
definiteness we will largely consider the approximately $AdS$ case in this paper, but will list more model
independent formulas in terms of $f$ and the potential $V$ for the density perturbation predictions in Appendix
B.

\para
The potential $V(\phi)$ in \eqn{act2} arises from the couplings of the D-brane to background RR fluxes,
supplemented by any couplings of the D-brane to degrees of freedom of the compactification (including quantum
generated effects). In the case of a pure $AdS_5\times X$ geometry, the potential is quartic.  In the case of
the ${\cal N}=4$ theory, this is $V=-\phi^4/\lambda$ which cancels the gravitational force, reflecting the BPS
nature of the D3-brane. In this case, conformal invariance prohibits the generation of a mass term for $\phi$.
However, for our approximate AdS throat, conformal invariance holds only at scales above $\phi_{IR}$ and there
is nothing preventing the generation of a mass for $\phi$, \be V(\phi)=m^2\phi^2 \label{m}\ee This leads to a
$\chi$ mass of order $\sqrt{g_s}m$ cutting off the throat, so the mass satisfies $m\leq \phi_{IR}/\sqrt{g_s}$.

\para
Because it is generic, we assume that such a mass is indeed generated, although it remains an important open
problem to explicitly calculate the scale $m$ in a controlled string compactification. In \cite{Dccel}\ it was
shown that this setup is self-consistent against back reaction of the probe (including the $m^2\phi^2$ energy)
on the throat spacetime, and against particle production effects including graviton emission into the
bulk.\footnote{There is also an interesting question as to whether there is a conformal coupling of the theory
\eqn{act1} and \eqn{act2}\ to four-dimensional gravity. Relevant calculations have been done in
\cite{Fotopoulos:2002wy}\ and \cite{Seiberg:1999xz}.  In our high velocity phase, the conformal coupling would
have velocity dependent effects.  In any case, in our inflationary phase, the conformal coupling is be a
subleading effect since the right hand side of the Friedmann equation is dominated by the potential energy, but
its presence would affect other cosmological solutions based on D-cceleration such as the dust phase described
in \cite{Dccel}.}

\para
While the above discussion has been phrased in terms of the five-dimensional geometry, there is an
interpretation of the action \eqn{act2} purely in terms of the holographic four-dimensional field theory with 't
Hooft parameter $\lambda$. Expanding the square-root, we see that this action contains an infinite number of
higher derivative terms. In the pure CFT, these are to be understood as arising from integrating out the $\chi$
modes from the microscopic field theory which become light as $\phi\rightarrow 0$. For example, in the simplest
case of global ${\cal N}=4$ super Yang-Mills, $\phi$ is a Higgs field breaking the $U(N)$ gauge group to
$U(N-1)$ and the $\chi$ fields are the Higgsed gauge bosons and their superpartners, each of which has mass
$m_\chi = \phi$ in the static vacuum. Indeed, in the case of ${\cal N}=4$ super Yang-Mills, one can prove that
the higher derivative terms arising from virtual W-bosons sum up to give the characteristic square-root form
{\it even} at weak coupling \cite{maldbigboy}. However, in the weakly coupled case, particle creation effects
must be taken
into account \cite{trappedinflation}, rendering the effective action ineffective for cosmological purposes. Note
that in the cutoff throat arising in the presence of our nontrivial mass term, we do not generically have a
large number of explicit stretched string $\chi$ modes.  However, model independently the throat contains bulk
gravity Kaluza Klein modes.

\subsubsection*{The Speed Limit and the Velocity-dependent Masses and Couplings}

The non-analytic behavior of the square-root in \eqn{act2} gives rise to a speed limit restricting how fast the
scalar field may roll. This, of course, is nothing but Einstein's causal speed limit in the holographic
dimension and is the crucial bit of new physics in our inflationary mechanism. When the AdS throat of equation
\eqn{f} provides a good approximation to the dual geometry, this speed limit is given by  $|\dot{\phi}| \leq
\phi^2/\sqrt{\lambda}$. A useful measure of how close we are to the limit is given by \be \gamma =
\frac{1}{\sqrt{1-f(\phi)\dot{\phi}^2}} \label{gamma}\ee which is analogous to the Lorentz contraction factor
defined in special relativity and grows without bound as the speed limit is approached.

\para
In the presence of large proper velocity, and thus large $\gamma$, the masses and couplings of modes in our
system are significantly altered from their static values.  For example, the mass of the $\chi$ modes becomes
\be m_\chi\sim {\phi\over\gamma}  \label{movingWmass}\ee while the decay rate of canonically normalized
perturbations of $\phi$ to each $\chi$ is also suppressed by powers of $\gamma$.  Although these features will
not play a large role in our analysis, they are an interesting aspect of the dynamics of the time dependent
field theory, so we derive these results in Appendix A.  Similarly we expect $\gamma$ dependence in the
conformal coupling when it is present, though as described above this effect is also subdominant in our
inflationary phase.

\para
The final velocity-dependent effect we wish to note here is one which will have observational importance in our
model. It is clear from the matter DBI action \eqn{act2}\ that expanding $\phi$ in fluctuations
$\phi\to\phi+\alpha$, which involves expanding the square root in \eqn{act2}, will produce powers of $\gamma$
accompanying the powers of the fluctuation $\alpha$. That is, the nonlinear terms in the perturbation $\alpha$
are enhanced by the $\gamma\gg 1$ D-cceleration mechanism.  We shall see in Section 4 that this simple
connection will lead to a lower bound on
the non-Gaussianity of the primordial density perturbations in a D-ccelerating system.

\subsection{Inflationary Cosmology from D-cceleration}

For $\gamma \gg 1$, the speed limit dominates the dynamics of the scalar field $\phi$ and leads to interesting
cosmologies which differ from those of the usual slowly rolling scalar field. We choose a flat FRW metric
ansatz, \be ds^2=-dt^2+a(t)^2\,\delta_{ij}dx^idx^j \nn\ee In \cite{Dccel}, the late time behaviour of the scale
factor $a$ and the scalar field $\phi$ were determined for a variety of potentials $V(\phi)$ and for
$f(\phi)=\lambda/\phi^4$ using the Hamilton-Jacobi approach. This method elevates the scalar field $\phi$ to the
role of cosmological time, so that the Hubble parameter $H=\dot{a}/a$ is considered as a function $H=H(\phi)$
determined in terms of the potential by,
\be
V(\phi)=3g_sM_p^2\,H(\phi)^2-\gamma(\phi)/f(\phi)
\label{V}\ee
where $\gamma(\phi)$ is given by
\be
\gamma(\phi)=\left(1+4g_s^2M_p^4\,f(\phi)\,H^\prime(\phi)^2\right)^{1/2}
\label{gamma2}\ee
The evolution of $\phi(t)$ is then fixed by the first order Friedmann equation,
\be
\dot{\phi}=-2g_sM_p^2\,\frac{H^\prime(\phi)}{\gamma(\phi)}
\label{phidot}\ee
It can be easily checked that when $\phi$ obeys its equation of motion, the definition of $\gamma(\phi)$ given
in \eqn{gamma2} coincides with the Lorentz contraction factor defined in \eqn{gamma}. As in the standard
inflationary scenarios, it will prove useful to introduce a slow-roll parameter. We choose \be \epsilon =
\frac{2g_sM_p^2}{\gamma}\left(\frac{H^\prime}{H}\right)^2 \label{slowrollepsilon}\ee which parameterises the
deviation from a pure de Sitter phase. In particular we have $\ddot{a}/a=H^2(1-\epsilon)$.

\para
In \cite{Dccel} the first order equations \eqn{V} and \eqn{phidot} were studied for a variety of potentials
$V(\phi)$. Here we shall focus on the massive scalar field \eqn{m}, for which the late time dynamics was shown
to be a power law inflation given by $a(t)\to a_0t^{1/\epsilon}$ and \be \phi \rightarrow
\frac{\sqrt{\lambda}}{t},\ \ \ \ \ \ \ \ \ \ \ \gamma \rightarrow \sqrt{\frac{4 g_s}{3\lambda}}\,M_p m\,t^2 ,\ \
\ \ \ \ \ \ \ \ \ \ \ H \rightarrow \frac{1}{\epsilon\,t} \label{wehavethepower}\ee The coefficient of the
(time-dependent) Hubble parameter is given by the slow-roll parameter which is a constant when evaluated on this
background, \be \frac{1}{\epsilon} = \frac{1}{3}\left( 1+\sqrt{1+\frac{3m^2\lambda}{g_{s}M_p^2}}\right) \approx
\sqrt{\frac{\lambda}{3g_{s}}}\frac{m}{M_p} \label{epsilon}\ee (Here the approximation is valid for $\epsilon\ll
1$).  For $\epsilon < 1$, we see that we have a phase of power-law inflation. The exponential de Sitter phase
can be thought of as the limit as $\epsilon\rightarrow 0$. Note that in contrast to usual single field
slow-roll inflation, we see that accelerated expansion occurs only if the mass of inflaton $m$ is suitably
large! This is one of several novel aspects of this model. In \cite{Dccel}, it was further shown that this
inflationary solution is robust against the expected further corrections to the action \eqn{act1}, including
higher derivative gravitational couplings and a conformal coupling if one exists in this setup (see
\cite{Fotopoulos:2002wy}\ for a discussion of this issue in a simpler setting).

\para
Although we will focus on the case $f=\lambda/\phi^4$ and $V=m^2\phi^2$, it is interesting to consider other
cases.  In Appendix B, we will collect more model-independent results for general $f$ and $V$.  In particular,
one finds that more generally the analogue of the slow roll conditions for us becomes \be
{V^\prime\over{V^{3/2}\sqrt{f}}}M_p\sqrt{g_s\over 3} << 1 \label{geneps}\ee and \be {{V^\prime}^2f\over
V}M_p^2g_s >> 1  \label{geneta}\ee These are to be contrasted with the usual slow roll conditions
$(V^\prime/V)^2M_p^2 << 1$ and $M_p^2V^{\prime\prime}/V << 1$.  Although these relations suggest inflation via
D-cceleration is enhanced by scaling up the potential energy $V$, it will turn out that the scalar power
spectrum scales like $V^2$ which is thus bounded by the data.

\subsubsection*{Parameters, E-Foldings and the Inflationary Scale}

Let us now list the parameters of our inflationary model. Much of this paper will be concerned with fixing these
parameters by matching to observation. The Lagrangian contains three basic dimensionless parameters which can be
tuned at will: the string coupling $g_s$, the 't Hooft coupling $\lambda = (R/l_s)^4$ (which is related to the
number of degrees of freedom in the field theory in a model dependent way related to the geometry of the
internal space $X$), and the hierarchy of mass scales $m/M_p$.
In addition, one can consider the possibility of
other scalar VEVs in the system (both static VEVs and more complicated time dependence of scalars other than
$\phi$ could come in).  This will partially be covered by our more model independent formulae in Appendix B in
terms of different warp factors.

\para
Given the above parameters determining the warped geometry, we also have the possibility to start and end
inflation anywhere within the AdS-like throat, subject to $\phi_{IR}\leq \phi_{end} < \phi_{start} \leq
\phi_{UV}$, where $\phi_{IR}$ and $\phi_{UV}$ are determined by the geometry of the throat.  For example, in
$AdS_5\times X$ geometries such as $AdS_5\times S^5$ geometries and those related by orbifolds of the $S^5$,
joined onto Calabi-Yau geometries in the manner of \cite{warped}, we have the following relations.  The UV end
of the throat arises at the scale $\phi_{UV} \alpha^\prime\sim R$. We can rewrite this in terms of the four
dimensional Planck mass $M_p$ by using the relation  $M_p^2\sim R \tilde V_X/(g_s^2l_s^8)$ where $\tilde V_X$ is
the volume in string units of $X$ (and we are assuming the throat contributes a leading contribution to the 4d
Planck mass). This gives
\be {\phi_{UV}\over \sqrt{g_s}}\sim {{\lambda^{1/8}M_p\sqrt{g_s}}\over\sqrt{\tilde V_X}}  \label{UVend}\ee
In order to avoid a functional fine tune of parameters, one requires $\phi/\sqrt{g_s}$ sufficiently small
relative to the Planck scale.  In matching to the data below, we will find that we require $\phi/\sqrt{g_s}$ to
start near the Planck scale to avoid overly large non-Gaussianity at late times.  It appears to be a model
dependent question whether $\phi_{start}$ begins somewhat above or somewhat below $M_p$; in the latter case this
would entail a functional fine tune of parameters as in early models of ordinary inflation.  In addition, we see
from \eqn{UVend}\ that this requires a sufficiently asymmetric compactification (i.e. a sufficiently small
$\tilde V_x$).

\para
Inflation could end either when the probe brane reaches the end of the throat at which point it undergoes
oscillations of the form discussed in \cite{kachru} or, more dramatically, if it annihilates with an anti-brane
at some point along its journey. From the above construction we impose that $\phi_{end} > \sqrt{g_s}m$ since in
the presence of our mass term the throat will close up at this level generically due to back reaction.
As with all brane-inflation models, reheating remains an important open problem: does the heat go into closed
string modes, or can it all be funneled into open strings? If the inflation occurs in a different warped
throat to the standard model, can the energy be transferred to standard model particles? For a review
os some of these issues, see \cite{Barnaby:2004kz}.


\para
We have already seen that inflation increases with increasing inflaton mass $m$. Specifically, we need
$\sqrt{\lambda}m
> \sqrt{g_s}M_p$. We can make this inequality more stringent by examining the number of e-foldings $N_e$ as
$\phi$ rolls along its inflationary path, \be N_e = \frac{1}{\epsilon}\,\log(\phi_{start}/\phi_{end})
\label{efoldings}\ee From the usual requirement of solving the flatness and homogeneity problems, we need at
least 60 e-foldings in total. Perhaps the simplest way to arrange for this is to set $N_e=60$ in our
inflationary window, but as far as current data is concerned, the 60 e-foldings could come from multiple passes
through the inflationary phase since the clean (nearly) scale invariant spectrum of the CMBR accounts for only
about 10 e-foldings of observations.  This may be natural in this setup, as the brane could ``bounce" back to
the top of the throat for multiple passes through our inflationary window, as in \cite{kachru}.  This
constraint, $N_e\ge 10$, gives one restriction on our parameters.


\para
Before computing the density perturbations, let us quickly examine the scale of inflation. Because we have a
time dependent Hubble parameter \eqn{wehavethepower} arising from power-law inflation, the energy scales at the
beginning and end of inflation differ and depend on the associated value of the inflaton. Using the bounds
above, we have \bea H_{\rm start} &\sim& \frac{1}{\sqrt{3g_s}}\frac{m\,\phi_{start}}{M_p}
\\
H_{\rm end}&\sim& \frac{1}{\sqrt{3g_s}}\frac{m\,\phi_{end}}{M_p}\ge \frac{1}{\sqrt{3}} \frac{m^2}{M_p} \ \ \ \ \ \
 \eea
>From which we see that the scale of inflation, and with it our chance to see tensor modes in the spectrum, is
tied to the mass $m$ of the inflaton. In the following, we shall see that we require a large inflaton mass $m$
(slightly below GUT scale) which favours observable tensor modes.

\section{Gaussian Perturbations}

In this section we compute the mode equations and solutions for the fluctuations of the inflaton and the metric
within the effective field theory \eqn{act1}. A priori, the analysis applies only for long wavelength fluctuations
of $\phi$, below the energies required to create $\chi$ particles (if they exist) and mix with the
KK modes in the full warped throat (the full strongly coupled field theory). In the case of the $U(N)$ ${\cal
N}=4$ SYM coupled to a compactification, this regime describes the low energy excitations on the approximate
Coulomb branch (on which the potential $V$ has been generated leading to time dependent solutions), where the
$U(N-1)$ and $U(1)$ sectors are approximately decoupled. With a slight abuse of language, we will refer to this
regime as the ``Coulomb branch" more generally.  In fact, we shall find that the mixing between
the inflaton and the rest of the system is negligible for the
scales of interest in our problem. This is in part because the couplings of the inflaton to the
10d bulk modes are Planck-suppressed, as well as because of the self consistency checks performed in
\cite{Dccel}. Therefore we will focus on the effective action for fluctuations of the inflaton $\phi$.

\para
As we shall see, an important scale arises because the fluctuations of $\phi$ do not propagate at the speed of light,
but rather
at a lower speed of sound $c_s$, resulting in an associated length scale $c_s/H$ we will refer to as the ``sound
horizon'', corresponding to a physical momentum scale $k/a=H/c_s$ below which the fluctuations are frozen out.

\para
In addition to the fluctuations of the inflaton $\phi$, our system has many other modes.  In the gravity side
description, these include Kaluza Klein modes in the bulk and $\chi$ modes, which are strings stretching from
the brane down the warped throat (which may or may not exist depending on the content of the IR end of the
throat). Even though these do not mix strongly with the inflaton, a priori, these non-inflatons could produce
isocurvature fluctuations that are too large. However, we will see that they can be consistently decoupled from
the low energy physics by requiring their masses to be much larger than $H$ (in the case of these modes, there
is no additional sound speed factor). We will derive the resulting constraint in section 3.2.

\para
The analysis of the linearized scalar fluctuations in models with non-standard kinetic terms was performed in
generality by Garriga and Mukhanov \cite{GM} and we follow their analysis closely. We start by decomposing the
scalar field into its rolling background value and a fluctuation $\alpha$, \be
\phi\rightarrow\phi(t)+\alpha(x,t) \ee with the scalar perturbation of the metric given in longitudinal gauge by
\be ds^2=-(1+2\Phi)dt^2+(1-2\Phi)a^2\,\delta_{ij}dx^idx^j \ee The metric perturbation $\Phi$ is simply the
Newtonian potential. The two fluctuations $\alpha$ and $\Phi$ are not independent: constraints in the
action leave only a single scalar degree of freedom, taken to be the linear combination \be \zeta =
\left(\frac{H}{\dot{\phi}}\right)\alpha+\Phi \ee It is this combination of fluctuations which becomes frozen as
it exits the horizon during inflation, later imprinting itself on the CMBR and seeding structure formation when
it re-enters. The evolution of $\zeta$ is determined by the linearized Einstein equations, but it is most simply
expressed if we first define a new variable \cite{GM}: $v\equiv \zeta z$ where \be z\equiv
\frac{a\gamma^{3/2}\dot{\phi}}{H}\ \rightarrow\ z_0 t^{(2\epsilon+1)/\epsilon} \nn\ee The second expression
holds on the late time inflationary background \eqn{wehavethepower} and $z_0$ is a constant involving the
various parameters  $m/M_p$, $\lambda$ and $g_s$. Finally we also introduce conformal time $d\tau = dt/a$ which,
as usual in an inflationary universe, runs from $\tau\in (-\infty, 0)$ as $t\in(0,+\infty)$. Decomposing into
Fourier modes $v=v_ke^{ikx}$, the equation of motion for $v$ is given by the fluctuation equation \be
\frac{d^2v_k}{d\tau^2}+\left(\frac{k^2}{\gamma^2}-\frac{1}{z}\frac{d^2z}{d^2\tau}\right)v_k=0 \label{v}\ee While
this determines the evolution of the fluctuations, to get the normalisation we also need the action; it suffices
to state that the kinetic term for $v/\sqrt{g_s}$ (in conformal time) is canonically normalised \cite{GM}.

\para
Before proceeding with the solutions to \eqn{v}, let us pause to point out the crucial fact that ripples in the
field $v$ do not travel at the speed of light, but at a lower sound speed $c_s$. To see this, consider large
wavenumber $k$ (still satisfying $k/a<\phi$ so this analysis applies), so that the second term in \eqn{v}
dominates over the third. In this limit we are left with a wave equation in flat space of the schematic form
$\ddot{v}+c_s^2v^{\prime\prime}$ where the speed of sound is given by \be c_s=\frac{1}{\gamma} \nn\ee For the
fluctuations $v$, the speed $c_s$ plays the role of the speed of light; for example, as we shall review, modes
of wavenumber $k$ freeze when they cross the {\it sound horizon} at $aH = c_sk$ rather than the true horizon at
$aH=k$ \cite{GM}. Comparing with the late time inflationary solution \eqn{wehavethepower}, we see that the speed
of sound is monotonically decreasing in our model $c_s\sim 1/t^2$.

\para
We now turn to a study of the solutions to \eqn{v}. We first need an expression for the
potential $(d^2z/d\tau^2)/z$ on our inflationary background \eqn{wehavethepower}.
Expanding in the slow-roll parameter $\epsilon$, have
\be
\frac{1}{z}\frac{d^2z}{d\tau^2}=2a^2H^2\left(1+\frac{5}{2}\,\epsilon\right)
\label{vequation}\ee
Re-expressing the Hubble scale in conformal time, $aH=1/(-1+\epsilon)\tau$, we have our
final expression for the fluctuations
\be
\frac{d^2v_k}{d\tau^2}+\left(c_s^2k^2-\frac{\nu^2-\ft14}{\tau^2}\right)v_k = 0
\label{itsawinner}\ee
where the level $\nu$ is given by
\be
\nu^2=\frac{2+5\epsilon}{(-1+\epsilon)^2}+\frac{1}{4}
\nn\ee
so from \eqn{efoldings} we see that for a large number of e-foldings so that
$\epsilon \approx 0$ we have
simply $\nu\approx 3/2$. Recall that the value $\nu=3/2$ is relevant for the
fluctuations in a pure de Sitter phase, as expected given that $\epsilon$
characterises the deviation from de Sitter.
>From \eqn{wehavethepower}, we see that $c_s=1/\gamma$ and $H$ change much more
slowly than the scale factor $a(t)$. Over suitably short time scales, we may
therefore approximate \eqn{itsawinner} as the equation for de Sitter fluctuations,
with the well-known Bessel function solution,
\be
\frac{v_k}{\sqrt{g_s}} = \ft12(-\pi\tau)^{1/2}\left[C_+H^1_{3/2}(-kc_s\tau)+C_-H^2_{3/2}(-kc_s\tau)\right]
\label{simplev}\ee
where $H^{(1,2)}$ are the Hankel functions, defined in terms of the usual Bessel functions $J$ and $N$ as
$H^{(1,2)}_\nu=J_\nu\pm iN_\nu$. The normalisation of the fluctuations is given by $|C_+|^2-|C_-|^2=1$.
Fluctuations arising from the standard Bunch-Davies vacuum state have $C_+=1$ with $C_-=0$ for which the early
time $c_sk\tau\rightarrow -\infty$ (or, equivalently, small wavelength) behaviour of the fluctuations contains
only Minkowski space positive frequency modes, \be H^1_{3/2}(-kc_s\tau)\ \rightarrow\ \sqrt{\frac{-2}{\pi k
c_s\tau}}\ e^{-ikc_s\tau} \nn\ee Previous authors \cite{transplanckian}
have argued that Planck scale physics may be encoded in
deviations from the Bunch-Davies vacuum to the ``$\alpha$ vacua" and further generalizations, although the
validity of such vacuum states remains controversial. For us the situation is a priori a little different: our
starting action \eqn{act2} is an effective theory, valid only above some length scale; even though fluctuations
of the microscopic theory may lie in the Bunch-Davies vacuum, this does not guarantee that the same is true of
the long-distance fluctuations $\zeta$. We shall examine this issue in more detail shortly. For now, let us
proceed with the analysis of the normalization of density fluctuations in an arbitrary vacuum specified by $C_+$
and $C_-$. At late times (or for large wavelengths) $c_sk\tau\rightarrow 0^{-}$, we have
%
%
%
%
%
%
%
%
\be
\frac{v_k}{\sqrt{g_s}}\ \rightarrow\  \frac{2i}{\tau}\,\frac{1}{(2c_sk)^{3/2}}\,(C_+-C_-)=-i(C_+-C_-)\,
\left(\frac{H^2}{\dot{\phi}}\right)\frac{z}{2^{1/2}k^{3/2}} \label{simplesoln}\ee where, in the second equality,
we have used $\tau\approx -1/aH$, to leading order in the slow-roll parameter. Recalling that $v_k=z\zeta_k$, we
rederive the standard result that $\zeta_k$ becomes constant at late times when it exits the horizon with
magnitude,
\be
\frac{\zeta_k}{\sqrt{g_s}} = i(C_+-C_-)\ \frac{1}{\epsilon^2\sqrt{\lambda}}\ \frac{1}{2^{1/2}k^{3/2}}
\label{approxzeta}\ee
The CMBR power spectrum is fixed in terms of these modes $\zeta$ which remain constant outside the horizon. It
is given by,
\be {\cal P}_k^\zeta = \frac{1}{2\pi^2}|\zeta_k|^2k^3 = \frac{1}{4\pi^2}|C_+-C_-|^2
\,\frac{g_s}{\epsilon^4\lambda}
\label{powerzeta}\ee
Note that the parametric scaling of the power spectrum can be
simply understood by the usual expression ${\cal P}^\zeta_k\sim g_sH^4/\dot{\phi}^2 \sim g_s/\epsilon^4\lambda$ where
we have used the inflationary background \eqn{wehavethepower} and the string coupling $g_s$ enters because
the canonically normalised scalar field is ${\phi}/{\sqrt{g_s}}$.

\subsection{The Vacuum}

Let us now discuss the issue of the vacuum, equivalently the coefficients $C_{\pm}(k)$ in the long distance mode
solutions. The linearized mode solutions are valid for modes on the Coulomb branch below the scale $\phi$ at
which it becomes kinematically possible to mix with the full CFT.  In the beginning of our inflationary window,
where the fluctuations seen now in the CMBR are generated, one can check that the corresponding length scale
$1/\phi$ is smaller than the sound horizon $\gamma/H$ at which the linearized mode solutions we found in the
last section freeze out.  As CFT modes stretch to become bigger than $1/\phi$, the dispersion
relation for $\alpha$ changes as its interactions with the rest of the CFT turn off.  A priori this provides a
possibility for particle production generating a nontrivial $C_-$ component to the vacuum even if we start in
the Bunch-Davies vacuum at short distances. However, because of the Planck suppression of couplings between the
brane and the bulk, we find in our situation that this mixing gives a negligible contribution to the primordial
density fluctuations. Therefore in the bulk of our analysis we will assume the adiabatic vacuum for the modes,
with $C_+=1, C_-=0$.  It is not difficult to extend this analysis to nontrivial vacua.

\subsection{Power Spectrum}

Here we will put together our results for various contributions to the CMBR power spectrum at the Gaussian
level, as functions of our parameters, leaving the computation of the non-Gaussian contributions to the next
section. Finally in section 5 we will put the constraints together to make numerical predictions for our
parameters. For the reasons discussed above, we work in the Bunch-Davies vacuum with $C_+=1$, $C_-=0$; the
generalization to other vacua is straightforward. The observed CMBR fluctuations correspond to the first 10 of
the last 60 e-foldings, so we will evaluate our results near the beginning of our inflationary window, subject
to the various constraints.

\para
Let us start with the contribution from scalar fluctuations. As discussed above, we find for the scalar
perturbations a power spectrum (in the Bunch Davies vacuum)
\be
{\cal P}^\zeta_k=\frac{1}{4\pi^2}\,\frac{g_s}{\epsilon^4\lambda} \label{bob}\ee

\subsubsection{The Scalar Tilt}

In principle, the spectrum of fluctuations may be tilted by three different effects. Firstly, as in usual
inflationary stories, the deviation of the space from de Sitter leads to a tilted spectrum. Secondly, the
shrinking sound horizon means that modes are freezing on ever smaller distance scales, again leading to a tilted
spectrum.  Finally, in the presence of a nontrivial vacuum, a third contribution to the tilt would come from the
$C_\pm(k)$. In fact, it turns out that the first two effects cancel each other out at order $\epsilon$, leaving
us with a tilt of order $\epsilon^2$ in the Bunch-Davies vacuum.

\para
To compute the tilt we must redo the calculation at the beginning of this section, no longer treating
$c_s$ as constant, but now including the varying sound speed.
Using the explicit temporal dependence $c_s=1/\gamma$ given
in \eqn{wehavethepower}, and working to leading order in $\epsilon$, the solutions to
\eqn{itsawinner} are given by, \be v_k=\ft12\sqrt{-\pi\tau}\left[C_+ H^1_{3/2}(\mu k\tau^{1+2\epsilon}) + C_-
H^2_{3/2}(\mu k\tau^{1+2\epsilon}) \right] \label{longun}\ee where the argument of the Hankel function includes
the factor $\mu = 3(1-2\epsilon)/2m^2\epsilon$. Notice in particular that the index of the Hankel function
remains $3/2$ even though we are working to leading order in $\epsilon$ (i.e. the index receives contributions
starting at $\epsilon^2$). This is the manifestation of the miracle mentioned above: the shrinking sound horizon
cancels the deviation from a de Sitter background to retain a scale invariant spectrum to order $\epsilon$. Thus
we obtain a tilt satisfying \be n_s-1= {\cal O}(\epsilon^2) \label{scalartilt}\ee

\subsubsection{The Tensor Modes}

The power ${\cal P}_k^h$ in tensor modes in our model is given by the same expression as in
standard inflationary models: ${\cal P}^h \sim (H/M_p)^2$ evaluated at
{\it causal} horizon exit $aH=k$. Including all the factors, we have at the end of inflation
\be {\cal P}_k^h = \frac{2}{3\pi^2}\frac{m^2\phi^2}{g_sM_p^4}
\label{tensorpower}\ee
The level of tensor modes in the CMBR are usually expressed as a fraction $r$ of the scalar
modes which, from \eqn{bob} and \eqn{tensorpower} takes the particularly simple form in
our model in terms of $\epsilon$ and $\gamma$,
\be
r=\frac{{\cal P}^h}{{\cal P}^\zeta} = \frac{16\epsilon}{\gamma}
\label{gammatensor}\ee
The tilt of the gravitational spectrum in our model can be simply computed \cite{GM} and
is given by
\be
n_T=\frac{d\ln{\cal P}^h_k}{d\ln k}= -2\epsilon
\nn\ee
As discussed in \cite{GM}, the relation between the tensor and scalar tilt is different from that in ordinary
weakly coupled slow-roll inflation.  In section 4 we will show that \eqn{gammatensor} leads to a simple relation
between the tensor power and the non-Gaussianity in the model.

\subsubsection{Isocurvature Perturbations}

Having determined the fluctuations due to the $\phi$ field and the graviton, let us now revisit the question of
the further non-inflaton degrees of freedom in the system. The lightest of these are the $\chi$ modes (when
present) and the Kaluza-Klein modes in the warped throat. The KK modes are decoupled from the inflaton $\phi$ at
long distances, and their masses and momenta are not suppressed by $\gamma$ factors and so they only freeze out
at the Hubble horizon (rather than at the smaller sound horizon at which the inflaton modes freeze out). The
$\chi$ modes also have an ordinary dispersion relation (with a suppressed mass \eqn{movingWmass}\ but
unsuppressed spatial momentum as can be seen from the nonabelian DBI action in Appendix A).   If any of these
particles are lighter than the Hubble scale $H$ during inflation, they may contribute isocurvature perturbations
which later processing can convert into curvature perturbations. Since there are many of these extra degrees of
freedom in our system, we must be particularly careful about their contributions.

\para
Let us first discuss the Kaluza-Klein modes in the warped throat. The KK masses are given by their UV value of
$1/R$ times the warp factor: \be m_{KK}={1\over R}{\phi_{IR}\over\phi_{UV}}. \label{KKmasses}\ee Requiring this
to be greater than $H$ leads to the constraint \be  m_{KK}>H \Rightarrow
{\phi_{UV}\over\phi_{IR}}\phi_{start}<\sqrt{3g_s}{M_p\over m}{1\over\lambda^{1/4}l_s} \label{KKdecoup}\ee The
relation between $l_s$ and the Planck scale $M_p$ depends on the internal volumes we have not specified
directly.  If the Calabi-Yau contribution to the Planck scale is subdominant to the warped throat contribution,
we obtain \be M_p^2\sim {R\over l_s^3}V_X \label{Planckreln}\ee where $V_X$ is the volume of the internal
component in the approximate $AdS_5\times X$ geometry of the warped throat.  Given this and the relation
$R/l_s\sim \lambda^{1/4}$ we can translate our condition \eqn{KKdecoup}\ into the condition \be
{\phi_{UV}\over\phi_{IR}}\phi_{start}<{\epsilon\lambda^{1/8}M_p\over \sqrt{V_X}} \label{KKfinal}\ee Let us now
discuss the $\chi$ particles.  In the original AdS solution, their mass satisfies \be {m_\chi\over H} =
{\phi\over{\gamma H}}\sim {\epsilon\sqrt{\lambda}\over\gamma} \label{chimassratio}\ee and we must require this
to be greater than one to avoid contributions to isocurvature perturbations from the $\chi$ particles.  As
discussed above, in the cutoff throat, these modes may no longer exist as they did in the infinite throat
describing AdS (it becomes a model-dependent question of where the strings could end in the IR region of the
throat).

\para
If the masses of the KK modes and $\chi$s are not taken above the Hubble scale, then their contributions may be
important, in a way that depends on less well understood aspects of their processing after inflation.  We will
run our analysis for both possibilities--keeping the extra modes decoupled and relaxing that restriction.

\section{Non-Gaussianity}

In this section we calculate the non-Gaussian corrections to the power spectrum. We start
by simply estimating the strength of the corrections,
after which we compute the full three-point functions including the momentum dependence.

\para
As before, let us consider the fluctuations of $\phi$ about the solution we have found. We perturb the solution
as
\be \phi\rightarrow \phi(t)+\alpha(x,t) \label{pe} \ee
Before computing the full three-point function, we start with an estimate of the strength of
interactions in the model relative to the Gaussian contribution. We do this, following
\cite{Creminelli:2003iq,Arkani-Hamed:2003uz}, by simply considering  the ratio of the cubic matter
Lagrangian to the quadratic matter Lagrangian evaluated on
the late time frozen out solution for $\alpha$, neglecting effects coming from mixing with gravitational
perturbations. This provides a rough estimate of the strength of the non-Gaussian
features, since it is the matter self-interactions which must dominate in order to obtain appreciable
non-Gaussianity.

\para
Naively since our field theory is strongly coupled, it is a priori in danger of producing from the matter sector
alone non-Gaussianity that is far too large.  In fact the strong 't Hooft coupling itself is not a harbinger of
large non-Gaussianity since there is a dual weakly coupled description on the gravity side--as we discussed in
the previous sections, the observed perturbations come from long distance modes stretched out by the inflation,
and at long distance the system is on its Coulomb branch where the inflaton is weakly coupled.  However, in the
moving system there is a danger of large $\gamma$ factors entering into the self-couplings of $\phi$ since
expanding the DBI action \eqn{act2} in perturbations $\alpha$ yields positive powers of $\gamma$.

\para
The action of the probe brane is given by \eqn{act2}. Therefore for the perturbation \eqn{pe} one has
\be
{\cal L}=-{a^3\over g_{\rm YM}^2}\left(\ {(\phi+\alpha)^4\over\lambda}
\sqrt{1-\frac{\lambda({\dot\phi}^2+2{\dot\phi}{\dot \alpha}+ {\dot
\alpha}^2-(\nabla\alpha)^2/a^2)}{(\phi+\alpha)^4}}+m^2(\phi+\alpha)^2\right)
\ee
which can be expanded in powers of the
fluctuations. At second order of fluctuations one gets
\bea
{\cal L}_2&=&{a^3\over g_{\rm YM}^2}\bigg{[}\frac{1}{2}{\gamma^3{\dot \alpha}^2} -
{\gamma\over 2a^2}(\nabla\alpha)^2-
{2\gamma^3\lambda{\dot\phi}^3\over\phi^5}\alpha{\dot\alpha}
\nn\\ &&\ \ \ \ \ \ \ \ \ \ \
+2\gamma^3\left( {\lambda{\dot \phi}^4\over
\phi^6}-{3{\dot\phi}^2\over 2\gamma^2\phi^2}- {3\phi^2\over\lambda\gamma^4}
-{m^2\over 2\gamma^3}\right)\alpha^2\bigg{]}
\eea
for which the equation of motion, to leading order in $\epsilon$ and $1/\gamma$, is
\be
{\ddot \alpha}+\left({6\over t}+3H\right){\dot \alpha}+ \left({6\over
t}+6H\right){\alpha\over t}-{\nabla^2\alpha\over a^2\gamma^2}=0
\ee
To third order in $\alpha$, one finds the Lagrangian
\bea
{\cal L}_3&=&{a^3\over g_{\rm YM}^2}\bigg{[}{\lambda\gamma^5 {\dot\phi}\over 2\phi^4}{\dot
\alpha}^3-{\lambda\gamma^3 {\dot\phi}^2\over a^2\phi^4}\left({{\dot\alpha}\over 2{\dot\phi}}-{\alpha\over
\phi}\right)(\nabla\alpha)^2 + {6\lambda^2\gamma^5{\dot\phi}^5\over \phi^{10}} \left(1+{5\phi^4\over
6\lambda\gamma^2{\dot\phi}^2}\right) {\dot\alpha}\alpha^2\cr &&\cr
&&\ \ \ \ \ \ \ \ -{3\lambda\gamma^5{\dot\phi}^2\over
\phi^5}{\dot\alpha}^2\alpha -4\gamma^5\left({\lambda^2{\dot\phi}^6\over \phi^{11}}+{{\dot\phi}^2\over
2\gamma^2\phi^3}+{\phi\over \lambda\gamma^6}
\right) \alpha^3
\bigg{]}\;.
\label{thatsmyboy}\eea
%
%
%
%
%
%
%
%
As derived in section 3, at the time they are formed the fluctuation modes in the Bunch-Davies vacuum scale like
\be \frac{\alpha}{\sqrt{g_s}}\sim H \ \ \ \ \ ,\ \ \ \ \ \ \frac{\dot\alpha}{\sqrt{g_s}}\sim H\alpha
\label{alphascale}\ee Plugging this into the expansions above, and recalling our expression for the Gaussian
power spectrum ${\cal P}_k^\zeta\sim g_sH^4/\dot{\phi}^2$ given in \eqn{powerzeta}, we obtain a parametric
estimate for the level of non-Gaussianity in our model, \be {{\cal L}_3\over {\cal L}_2}\sim
{H^2\over\dot\phi}\sqrt{g_s}\gamma^2  \sim \sqrt{{\cal P}_k^\zeta}\,\gamma^2 \label{nonG} \ee This simple result
ties the strength of the D-cceleration effect (as measured by $\gamma$) to the strength of the
non-Gaussianities. The WMAP limitations on the latter, roughly corresponding to ${\cal L}_3/{\cal L}_2\le
10^{-2}$ \cite{Acquaviva:2002ud}, then lead to an {\it upper} bound on $\gamma$.  There being also a lower bound
on $\gamma$ in order to accomplish acceleration on a steep potential through our D-cceleration mechanism, we
find ourselves in a very predictive situation -- our model based on the D-cceleration mechanism requires
relatively strong non-Gaussianity and is therefore observationally falsifiable.

\para
Note that for later times $\gamma$ increases and the nonlinear effects in our model grow.  This may have
implications for other aspects of structure formation in the model.

\para
Let us now proceed to compute the full three-point function. Not only will this present us with the correct
normalisation of this scaling, but it will also give us the important momentum dependence of the
non-Gaussianity. As in the cases discussed in  \cite{Creminelli:2003iq,Arkani-Hamed:2003uz}, the calculation of
large, unsuppressed non-Gaussian effects in the CMB is somewhat simpler than the corresponding calculation in
standard slow-roll inflation \cite{juan} since outside the horizon one finds that the effects of the matter
self-interaction dominate over the subtle corrections to the dictionary between $\zeta$ and $\alpha$ derived in
\cite{juan}. This allows us to straightforwardly compute $\langle\alpha^3\rangle$ and subsequently translate to
$\langle\zeta^3\rangle$ which determines the non-Gaussianity. In our large $\gamma$, small $\epsilon$ regime,
the leading order terms in \eqn{thatsmyboy} give us the interaction Hamiltonian,
\be {\cal H}_{\rm
int}=-\frac{a^3\gamma^5}{g_s}\left(
\frac{1}{2\dot{\phi}}\dot{\alpha}^3+\frac{k^2}{a^2\gamma^2}\frac{1}{2\dot{\phi}}\dot{\alpha}\alpha^2\right) \ee
which we use to compute the three-point function to first order,
\be
\langle\alpha_{k_1}(t)\alpha_{k_2}(t)\alpha_{k_3}(t)\rangle = -i\int_{t}^{t^\prime} dt^\prime\,
\langle\,[\,\alpha_{k_1}(t)\alpha_{k_2}(t)\alpha_{k_3}(t)\, ,\, {\cal H}_{\rm int}(t^\prime)\,]\,\rangle \ee
If
we decompose the fluctuations $\alpha$ in the usual fashion using creation and annihilation operators,
$\alpha_k(t)=u_k(t)a_k+u^\star(t)a_k^\dagger$, with canonical commutation relations
$[a_k,a^\dagger_{k^\prime}]=\delta_{k,k^\prime}$, then we find, \bea
\langle\alpha_{k_1}\alpha_{k_2}\alpha_{k_3}\rangle &=& -\frac{i}{g_s}(2\pi)^3\delta^{(3)}\left(\sum
\vec{k}_i\right) u_{k_1}(0)\,u_{k_2}(0)\,u_{k_3}(0)\,
\int_{-\infty}^0\frac{d\tau}{H\tau}\,\frac{\gamma^5}{2\dot{\phi}} \label{ouch}\\ && \ \ \ \
\cdot\left(6\frac{du_{k_1}^{\star}}{d\tau}\frac{du_{k_2}^{\star}}{d\tau} \frac{du_{k_3}^{\star}}{d\tau}
+\frac{2}{\gamma^2}\vec{k}_1\cdot\vec{k}_2\,u_{k_1}^\star(\tau) u_{k_2}^\star(\tau)\frac{du_{k_3}^\star}{d\tau}
+ {\rm sym} \right) +{\rm c.c.} \nn\eea where $+\ {\rm sym}$ permutes the three momenta $k_i$ in the term with
$\vec{k_1}\cdot\vec{k_2}$. From section 3, we have the solution for $u_k$,
\be
\frac{u_k(\tau)}{\sqrt{g_s}}=-\frac{\sqrt{-\pi\tau}}{2\gamma^{3/2}}\tau H\,H_{3/2}^{(1)}(-k\tau/\gamma)
=-\frac{iH}{\sqrt{2k^3}}\,e^{-ik\tau/\gamma}\, (i-k\tau/\gamma) \ee
Substituting this into the three-point
function \eqn{ouch} and translating to the variable $\zeta$ which remains constant outside the horizon, we find
our final expression for the three-point function including the angular dependence,
\bea
\langle\zeta_{k_1}\zeta_{k_2}\zeta_{k_3}\rangle &=& -\frac{ig_s^2}{2^3}(2\pi)^3\delta^{(3)}\left(\sum
\vec{k}_i\right) \left.\frac{H^6}{\dot{\phi}^3}\right|_{\tau=0}\, \prod_i \frac{1}{k_i^3} \
\int_{-\infty}^0\,d\tau\, \frac{H^2}{2\dot{\phi}}\, e^{i\sum_ik_i\tau/\gamma} \label{3pt}\\ &&
\cdot\left(\frac{6\tau^2}{\gamma}k_1^2k_2^2k_3^2-2\gamma
\left(i+\frac{k_1\tau}{\gamma}\right)\left(i+\frac{k_2\tau}{\gamma}\right) (\vec{k_1}\cdot\vec{k_2})\,k_3^2 +
{\rm sym}\right) + {\rm c.c} \nn\eea
As discussed in \cite{juan,Creminelli:2003iq,Arkani-Hamed:2003uz}, the
observational constraints \cite{Komatsu:2003fd}\ are quoted in terms of a quantity $f_{NL}$ defined by the
assumption of non-Gaussianity emerging from a nonlinear relation $\zeta =
\zeta_g-\ft35f_{NL}(\zeta_g^2-\langle\zeta_g^2\rangle)$ between the Gaussian curvature perturbation $\zeta_g$
and $\zeta$, the corrected curvature perturbation good at next to leading order. This assumption is not a
faithful representation of the three-point functions emerging from the self-interactions of inflation models,
leading instead to an angular dependence of the form\footnote{Here we use the WMAP convention for $f_{NL}$
which differs by a sign from that defined in \cite{juan}.},
\be \langle\zeta_{k_1}\zeta_{k_2}\zeta_{k_3}\rangle =
(2\pi)^7\delta^{(3)}\left(\sum \vec{k}_i\right)\left(-\ft35 f_{NL}({\cal
P}^\zeta)^2\right)\frac{4\sum_ik_i^3}{\prod_i2k_i^3} \label{fng}\ee
Nevertheless, the magnitude of $f_{NL}$ can
be determined by comparing \eqn{3pt} and \eqn{fng} evaluated on an equilateral triangle, $k_1=k_2=k_3$.
Performing the integral \eqn{3pt} under the assumption of the slow-roll condition, on a contour which leads to
suitable convergence, we find our non-Gaussianities are characterized by \be f_{NL}\approx (-0.32) \gamma^2
\label{lastone}\ee The magnitude of the non-Gaussianity is thus bounded from below in our model.  The
D-cceleration mechanism requires $\gamma$ at least somewhat greater than  $1$; since we have expanded in powers
of $1/\gamma$ we require for control, say, $\gamma>5$.  A lower bound on $\gamma$ would also follow from an
upper bound on $\phi$, since we can express $\gamma$ as \be \gamma\sim {2g_s\over\epsilon}{M_p^2\over \phi^2}.
\label{gammaandphi}\ee This will come into our analysis of the constraints from data, and lead us to a regime of
approximately Planckian field VEVs. Finally, recall from the computation of tensor modes $r=16\epsilon/\gamma$,
ensuring that an upper bound on non-Gaussianity (and therefore on $\gamma$) from the data leads to a lower bound
on the tensor power.

\section{Parameter Values and Predictions}

Having derived expressions for the various components of the CMBR, let us now compare with observational bounds.
The WMAP analysis \cite{Peiris:2003ff}\ gives, \be {\cal P}^\zeta \approx 2.95\times 10^{-9}A(k_0)\ \ \ \ ,\ \ \
r={\cal P}^h/{\cal P}^\zeta < r(k_0) \nn\ee where the observables are defined at $k=k_0=0.002\ Mpc^{-1}$. The
precise values of $A(k_0)$ and $r(k_0)$ depending on the data sets and priors included are provided in Table 1
of \cite{Peiris:2003ff}; the preferred values for the ratio $r$ vary from $0.29$ to $1.28$ and those of $A$
range from $0.71$ to $0.75$. For the tilt $n_s=d \ln {\cal P^\zeta}/d \ln k$, WMAP obtains central values
ranging from $1.13$ to $1.20$. Including data from the SDSS galaxy survey, WMAP and other sources, the preferred
value for the tilt is $0.97$, while the tensor modes are constrained by $r<0.5$ at $95\%$ CFL \cite{tegmark}.

\para
For the non-Gaussianity, the WMAP analysis finds a spectrum consistent with Gaussian primordial fluctuations
with $|f_{NL}|\leq 100$ while the Planck satellite will improve this bound to $|f_{NL}|\leq 5$. Other
statistical tests of the WMAP data have reported deviations from a Gaussian spectrum \cite{nong}, although we
here assume the more conservative WMAP analysis. From \eqn{lastone}, we find a constraint on the speed limit
due to the bound on non-Gaussianities
\be \gamma \leq 17
\label{gng}\ee
This translates into a lower bound on
$\epsilon$, determined by the value of the inflaton when the CMB perturbations are created,
\be \epsilon >
\frac{1}{10}\left(\frac{M_p^2g_s}{\phi^2}\right) \label{ep}\ee
Note that this requires that the value of the
canonically normalised inflaton field $\phi/\sqrt{g_s}$ cannot be far beneath the Planck scale $M_p$ for our
inflationary mechanism to hold. In fact, it will turn out that to keep non-Gaussianity suppressed throughout the
10 e-foldings of inflation that can be seen in the CMBR, our simple model requires an inflaton vev slightly
above the Planck scale. This can be avoided in the more general models described in Appendix B.

\para
Using our expression for the tensor power spectrum $r=16\epsilon/\gamma$, we see that for a given $\epsilon$,
the non-Gaussianity bound \eqn{gng} provides a {\it lower} bound on the spectrum of tensor perturbations. For
example, if we suppose that $\phi=\sqrt{g_s}M_p$, so that $\epsilon>1/10$, then the lower bound on the tensor
modes is $r>1/10$, a level which can in principle be detected by future dedicated polarization
experiments such as \cite{church}.

\para
We now give examples of our parameters at the two extremes of the slow-roll regime. From \eqn{ep}, the smallest
$\epsilon$ consistent with a Planck scale inflaton vev is (up to a factor of 2)
 $\epsilon = 1/20$. Then the COBE normalisation
requires an inflaton mass $m=2\times 10^{-5} M_p\approx 4\times 10^{13}$ GeV. The number of efoldings
\eqn{efoldings} arising in this model is $N_e < 200$. For these values, one finds the minimum proportion of
tensor modes $r=1/50$, and the maximum non-Gaussianity $f_{NL}\sim 100$. In fact, this non-Gaussianity holds
only at the beginning of inflation and grows with time, so that smaller scales on the CMB would already exhibit
too much deviation from a Gaussian distribution to be compatible with the WMAP analysis \cite{Komatsu:2003fd}.
This can be avoided by starting the inflaton slightly above the Planck scale ($\phi\sim 1.5\sqrt{g_s} M_p$ will
suffice). Alternatively, one may be able to bring down the non-Gaussianity by considering more general models
such as those with different warp factors $f(\phi)$ described in Appendix B.

\para
We can also examine the consequences of a larger slow-roll parameter, say $\epsilon \sim 1/5$. (Of course, we
still require the $\epsilon$ expansion to be valid so $\epsilon$ cannot be much larger than this). Then the COBE
normalisation sets $m\sim 10^{-4}M_p \sim 10^{14}$ GeV and we get at most 46 e-foldings. Subsequent e-foldings,
required to solve the flatness and homogeneity problems but require further fine tuning. Alternatively, one
could have piecewise periods of inflation, generated by another pass through the
inflationary window, as described in section 2, or by another mechanism. For sub-Planckian inflaton, the
non-Gaussianity starts at $f_{NL}\sim 6$ and increases to $f_{NL}\sim 100$ within 4 e-foldings, while the tensor
modes are now much higher, starting at low multipoles with $r\sim 8/25$. Again, this non-Gaussianity can be
lowered by starting slightly above the Planck scale, or may be mitigated by considering other warped
backgrounds.

\para
In addition to satisfying the CMBR constraints, we must consider the effects of the KK and $\chi$ modes. One
possibility is to completely decouple these by insisting on masses bigger than H. Using the requirements of
section 3.2.3, we find both of these modes can be decoupled in our inflationary phase.  In the case of the KK
modes, this requires ending the throat at a value of $\phi_{IR}\gg m\sqrt{g}$ and by making use of the freedom
to tune $g_s$ and $V_X$ (small tunings are required in both cases).

\subsection{Tuning Tricks and Tests}

Above we expressed our results in terms of the slow roll parameter $\epsilon$ and the mass ratio $m/M_p$, In
each case, this translates into an enormous value for $\lambda$. For example, when $\epsilon=1/20$, we have
$\lambda/g_s\sim 10^{12}$. If we instead take $\epsilon =1/5$, this is reduced to $\lambda/g_s\sim 10^{10}$. In
terms of the gravity dual $\lambda\sim (R/l_s)^4$, requiring that the radius of the AdS space $R$ is of order
$100-1000$ times the string length $l_s$.  The weakly coupled description of the system is the gravity side, so
this is the most direct way to express the tuning we require.  (In addition, as we have noted above, we are
forced by the growth of non-Gaussianities to consider a starting field VEV slightly above the Planck scale;
depending on the details of the model this may require a functional fine tune of parameters.)

\para
As far as the tuning of $\lambda/g_s$ goes, given the holographic duality involved it is interesting to
translate this into the dual field theoretic description related to the worldvolume theory on the branes making
up the throat. The tuning translates in a model-dependent way into a corresponding statement about the degrees
of freedom of the dual field theory. In (a deformation of) the $U(N)$ ${\cal N}=4$ SYM theory, we would have
$\lambda\sim g_s N$ and thus an enormous gauge group corresponding to $N=\lambda/g_s$ D3-branes. However by
placing $N$ 3-branes at an orbifold point, one can easily reduce this number.  For example, $N$ D3-branes at a
$Z_k$ orbifold point corresponds to $\lambda\sim g_s N k$ and by taking $k=N$ we reduce $N$ by 5 orders of
magnitude. Similarly, we may take a $Z_n^l$ orbifold group yielding $\lambda\sim g_sn^l N$ for which no integer
quantum number need be particularly large.  (This is reminiscent of the Bousso-Polchinski mechanism for making
the large number involved in tuning the cosmological constant from small input flux and topological quantum
numbers \cite{BP}.) These reductions correspond in the gravity side description to reducing the size of the
internal space $X$ in our approximate $AdS_5\times X$ solution.

\para
There is a lower bound on the number of degrees of freedom, if we identify this number with the holographic
Susskind-Witten entropy \cite{susswitt}\ in the throat given by \be S \sim R^3 M_5^3 \sim {R^3 V_X\over {g_s^2
l_s^8}} \label{SWentropy}\ee so if we take at minimum $V_X/l_s^5$ and $g_s$ to be of order 1, we obtain an
entropy \be S \ge \left({R\over l_s}\right)^3 \sim \lambda^{3/4}. \label{stillprettybad}\ee

\para
The tuning we need here is perhaps reminiscent of that arising in the first models of ordinary slow roll
inflation where the coupling had to be tiny (e.g. $\lambda_4\phi^4\sim 10^{-14}\phi^4$) rather than
huge as in our case.
Of course in ordinary slow roll inflation, later models with much less tuning
have been developed (including hybrid inflation \cite{Linde:1993cn}\ and ``new old" inflation
\cite{Dvali:2003vv}). We can hope that such generalizations will arise also in the case of our mechanism, and in
Appendix B we take the first steps toward a more model independent formulation.
%
%
%
%
%

\section{Conclusions and Future Directions}

We have shown in this paper that the D-cceleration model of inflation introduced in \cite{Dccel} is
observationally testable, primarily through a rather distinctive non-Gaussian contribution to the power spectrum
whose magnitude cannot be reduced below a strongly observable lower bound in this family of models. If the
predicted strong non-Gaussian signal is not seen by the Planck satellite, it will rule out this family of
models. Moreover, we strongly favor an observable spectrum of tensor fluctuations. Since models with large
primordial non-Gaussianity are rare (with the intriguing case of ``ghost inflation" \cite{Arkani-Hamed:2003uz}\
the only other proposal inhabiting this region), an observation of non-Gaussianity would raise the exciting
possibility of observational detection of a strongly 'tHooft coupled hidden sector.

\para
There are many postinflationary dynamical processes to be worked out in our model, such as reheating, formation
of cosmic strings, isocurvature perturbations in the case with light KK and $\chi$ modes, and the effects of our
strongly nonlinear evolution at very late times (which may be similar to ``tachyon matter"
\cite{Frolov:2002rr}\cite{Felder:2002sv}).

\para
It is also very much of interest to consider more systematically the space of models realizing the D-cceleration
mechanism.  One starting point for that is the set of formulas collected in Appendix B for more general warp
factors coming from other Coulomb branch configurations.  Rotating and thermal solutions provide explicit
examples of other generalizations of interest.

\para
There are also broader applications of the mechanism.  As in the related work \cite{trappedinflation}, the
inflationary phase here could also arise later in the evolution of the universe, diluting unwanted relics for
example; the D-cceleration phases more generally also dynamically leads the theory to get stuck near more
symmetric configurations.

\para
On a more formal level, the D-cceleration mechanism suggests the following interpretation of horizon physics
more generally.  Namely, we have seen \cite{Dccel}\ that the fact that the probe takes forever to
reach the AdS Poincare patch horizon is due to its interaction with the $\chi$ modes becoming light at the
horizon. Much more generally, general relativity predicts that a static observer in a spacetime with horizon
(such as de Sitter space and black hole and brane solutions) sees a probe take forever to reach the horizon. The
DBI action of the probes (accessible in concrete string models of such backgrounds) which describes this
slowdown rather generally contains contributions scaling like those obtained by integrating out $\chi$ modes
becoming light as the probe approaches the horizon.  These modes are closely related to the horizon entropy, so
we obtain in this way an interpretation of the GR-predicted probe slowdown in terms of interaction with the
microscopic degrees of freedom related to the horizon entropy.

\section*{Acknowledgements}
We would like to thank N. Arkani-Hamed, E. Baltz, T. Banks, E. Copeland, P. Creminelli, G. Dvali, R. Easther, A.
Guth, S. Kachru, L. Kofman, A. Linde, L. McAllister, S. Rahvar, A. Riotto,  H. Reall, S. Shenker, M. Van
Raamsdonk and T. Wiseman for useful discussions on this and related issues. We would like to thank Xingang Chen
for two corrections.  M.~A. is supported in part by Iranian TWAS chapter based at ISMO. E.~S. is supported in
part by the DOE under contract DE-AC03-76SF00515 and by the NSF under contract 9870115. D.~T. is supported by a
Pappalardo fellowship and is grateful to Neil Pappalardo for his generosity. D.T. would also like to thank SLAC
and the Department of Physics, Stanford University for their very kind hospitality. This work was also supported
in part by funds provided by the U.S. Department of Energy (D.O.E.) under cooperative research agreement
\#DF-FC02-94ER40818.

\section*{Appendix A: The Velocity-Dependent $\chi$ Mass}

In backgrounds with $\dot{\phi}\neq 0$, there are velocity dependent corrections to the effective action. For
example, as we mentioned in Section 2, the effective mass is shifted from $\phi$ in the static system to
\be m_\chi\sim {\phi\over\gamma}
\label{shiftmchi}\ee
This played a role in Section 3.1 as it determines the scale at which the microscopic theory kinematically is to
be matched onto the long distance effective theory for $\phi$.
%
%
We will study in our warped throat the mass of
a string stretched from our probe brane to another brane much closer to the horizon, as a function of the
velocities of the branes. As described in \cite{Tseytlin:1997cs}\cite{Taylor:1999pr}, the action for multiple
D3-branes in a background metric field with negligible covariant derivatives of field strengths is given by
(ignoring the potential terms from RR flux)
\bea S_{DBI} &=& -\int d^4x STr\left[\sqrt{det(\delta_i^j+i\alpha^{\prime\,-1}g_{ik}[X^k,X^j])}\right.
\label{nonabelianDBI} \\ && \ \ \ \ \ \ \ \ \ \ \ \ \ \ \ \ \ \left.
\sqrt{det\left(g_{\mu\nu}+D_\mu X^k(g_{kl}+i\alpha^{\prime\,-1}g_{kq}[X^q,X^r]g_{rl})D_\nu
X^l\right)}\right]\;. \nn \eea
Here the $i,j$ indices label directions transverse to the branes and $\mu,\nu$ indices label the directions
along the branes.  The $X^i$ are matrices describing the adjoint scalar fields on the collection of branes, and
the metric $g_{MN}$ is a function of these matrices.  The determinants in \eqn{nonabelianDBI}\ are taken over
the Lorentz indices.  The gauge indices are traced over -- this part of the action (which ignores commutators of
$F$ and $D_M X$) involves a ``symmetrized trace" $(STr)$ which requires summing over all orderings of the
adjoint matrices and then tracing over the gauge indices.

\para
The basic effect that reduces the $\chi$ mass in the moving system can be seen in a relatively simple way from
\eqn{nonabelianDBI}\ as follows.  The $\chi$ mass comes from a commutator of the background scalar fields
describing the motion of the pair of branes in one direction (which can be encoded in a diagonal matrix in an
appropriate basis) and the $\chi$ excitations in another direction (contained in an off-diagonal matrix in this
basis).  Expanding about the moving brane configuration, the first square root factor in \eqn{nonabelianDBI},
which contains a $\Delta \phi^2\chi^2$ mass term, is multiplied by $1/\gamma$ from the second square root
factor. The $\dot\chi^2$ term comes from expanding the second square root factor, which leads to its being
multiplied by one power of $\gamma$.  The net effect of this is a $\chi$ mass scaling like $\Delta\phi/\gamma$.

\para
This depended on no contribution to the effective $\chi$ mass term coming from the expansion of the second
square root in \eqn{nonabelianDBI}.  This follows from the symmetry of the expression with respect to its
Lorentz indices, as follows.  A contribution to the $\chi$ mass from the second square root factor would have to
arise from a contraction $\dot X^1 g_{1q}[X^q,X^l]g_{l1}\dot X^1$ where
$X^1={\rm diag}(\phi_1\alpha^\prime,\phi_2\alpha^\prime)$ is the adjoint scalar matrix describing the background
motion in the 1 direction.  But this contribution cancels by symmetry.

\para
Let us now discuss this analysis in more detail for our case. In the $AdS_5\times S^5$ throat in which the
branes propagate, the metric is
\be ds^2={\phi^2\alpha^\prime\over\sqrt{\lambda}}(-dt^2+d\vec
x^2)+\alpha^\prime\sqrt{\lambda}{d\phi^2\over\phi^2}+\alpha^\prime\sqrt{\lambda}d\Omega_5^2 \label{AdSmetric}
\ee
where $d\Omega_5^2$ is the $S^5$ metric (more generally we could have a different internal space) and the AdS
radius is $R=\sqrt{\alpha^\prime}\lambda^{1/4}$.
We are considering motion in the radial direction $\phi$:
\be \Phi = \left(
\begin {array}{cc}
\phi_1 &0\\
\noalign{\medskip} 0&\phi_2
\end {array}
\right) \label{phimatrix} \ee
where $\phi_1$ is the position of the probe brane and $\phi_2$ is the position of the brane on which the $\chi$
string is ending (which we will take to satisfy $\phi_2\ll\phi_1$). We would like to also consider fluctuations
in the transverse directions off diagonal in the gauge indices. For simplicity let us take
$d\Omega_5^2=d\theta^2+\dots$ and consider the corresponding $\Theta$ adjoint matrix:
\be {R\over\alpha^\prime}\Theta = \left(
\begin {array}{cc}
0&\chi \\
\noalign{\medskip} \chi &0
\end {array}
\right) \label{chimatrix} \ee
Plugging these matrices into \eqn{nonabelianDBI}\ and taking the traces yields the action
\bea S_{DBI} &=& -\int d^4x \left[
{\phi_1^4\over\lambda}\sqrt{1-{\lambda\dot\phi_1^2\over\phi_1^4}-{\sqrt{\lambda}\dot\chi^2
\alpha^\prime\over\phi_1^2}}\;\sqrt{1+(\Delta\phi)^2\chi^2{\sqrt{\lambda}\alpha^\prime \over\phi_1^2}} \right.
\label{tracedaction} \label{bigDBI} \\ && \ \ \ \ \ \ \ \ \ \ \ \ \ +\ \left.
{\phi_2^4\over\lambda}\sqrt{1-{\lambda\dot\phi_2^2\over\phi_2^4}-{\sqrt{\lambda}\dot\chi^2
\alpha^\prime\over\phi_2^2}}\; \sqrt{1+(\Delta\phi)^2\chi^2{\sqrt{\lambda}\alpha^\prime \over\phi_2^2}}\right]\;.
\nn\eea
Note that the second term is suppressed by powers of the warp factor at the position $\phi_2$ of the second
brane.  Indeed, expanding out \eqn{tracedaction}\ we find
\be m_\chi^2={\Delta\phi\over \gamma_1^2}\left[ {{1+{\phi_2^2\gamma_1}/{\phi_1^2\gamma_2}}\over
{1+{\phi_2^2\gamma_1}/{\phi_1^2\gamma_2}}}\right]\to {\phi_1^2\over \gamma_1^2}~~{\rm as}~\phi_2\to 0
\label{massDBI}\ee
This result agrees nicely with the spirit of \cite{KL}, where it was suggested that the limiting velocity should
be reflected in some modes becoming massless.  Indeed, we see that the $\chi$s become
massless as $\gamma\to\infty$.


\section*{Appendix B:  More General Models:  Choose your own adventure}

In the bulk of this paper, we considered a model of D-cceleration with a near horizon geometry given
approximately by $f(\phi)=\lambda/\phi^4$ in \eqn{act2}\ (corresponding to approximate AdS geometry), and with a
(generically generated) $m^2\phi^2$ potential.  In this appendix we will collect results for more general
$f(\phi)$ and $V$, which can come from a variety of different field theories coupled in different ways to
compactification sectors.  For example, in the case of the ${\cal N}=4$ SYM theory we could consider a probe
brane moving the presence of the $N-1$ other branes not all sitting statically at the origin. The adjoint scalar
eigenvalues corresponding to the other brane positions could be nonzero (and nonstatic in general).  From these
degrees of freedom we have not found models qualitatively different in predictions from that described in the
main paper, but we have not performed a complete systematic search for such models.

\para
Let us write the conditions for acceleration ($\ddot a>0$) and D-cceleration ($\gamma>1$) in terms of a general
$f$ and $V$.  In order to have acceleration, the potential energy must dominate over the kinetic energy in the
Friedmann equation, with the kinetic energy $\gamma/f >>1/f$ for D-cceleration.  We can formulate this condition
as follows. When satisfied, it means $H^2\sim {1\over {3 g_s M_p^2}}V$ and thus \be H^\prime \sim {1\over
2\sqrt{3g_s}M_p}{V^\prime\over V^{1/2}}\label{hprime}\ee Thus from \eqn{V}\ we obtain that the condition that
$V$ dominates is \be {V^{3/2}\over V^\prime M_p}\sqrt{{3f\over g_s}} \gg 1\label{Vdomination}\ee Similarly from
\eqn{gamma2}\ we find that for D-cceleration ($\gamma\gg 1$) we require \be {g_s\over 3V}{V^\prime}^2f M_p^2 \gg
1 \label{Dccelgen}\ee The number of e-foldings is \be N_e=\int_{\phi_f}^{\phi_i}{{d\phi}\over M_p}
\sqrt{{fV\over{3g_s}}} \label{genefoldings}\ee These conditions should be compared to the usual slow roll
conditions.

\para
Now given that $\gamma\gg 1$, we see that \be \dot\phi^2\sim {1\over f} \label{genphidot}\ee So the scalar power
spectrum ${\cal P}\sim {1\over 4\pi^2}|C_+-C_-|^2{H^4g_s\over \dot\phi^2}$ becomes in terms of $f$ and $V$ \be
{\cal P}^\zeta_k = {1\over {4 \pi^2}}|C_+-C_-|^2{fV^2\over{9g_sM_p^4}} \label{genscalarpower}\ee

\para
It is also possible to work out similarly the effects of a different warp factor on the level of non-Gaussianity
and the range of $\phi$.  Our preliminary analysis indicates that the same level of non-Gaussianity arises more
generally, but that different warp factors can adjust the bounds on $\phi_{start}$ somewhat and can lead to a
larger tilt.

\end{document}